\documentclass[aps,prb,twocolumn,floats]{revtex4}

\usepackage{amssymb}
\usepackage{amsmath}
\usepackage{graphicx}
\usepackage{bm}
\usepackage{mdframed}
\usepackage[pangram]{blindtext}

\begin{document}

\bibliographystyle{prsty}
\author{Hector Ochoa,$^1$ Se Kwon Kim,$^1$ Oleg Tchernyshyov,$^2$ and Yaroslav Tserkovnyak$^1$}
\affiliation{$^1$Department of Physics and Astronomy, University of California, Los Angeles, California 90095, USA \\
$^2$Department of Physics and Astronomy, Johns Hopkins University, Baltimore, Maryland 21218, USA}

\begin{abstract}
We theoretically study the dynamics of skyrmion crystals in electrically-insulating chiral magnets subjected to current-induced spin torques by adjacent metallic layers. We develop an elasticity theory that accounts for the gyrotropic force engendered by the non-trivial topology of the spin texture, tensions at the boundaries due to the exchange of linear and spin angular momentum with the metallic reservoirs, and dissipation in the bulk of the film. A steady translation of the skyrmion crystal is triggered by the current-induced tensions and subsequently sustained by dissipative forces, generating an electromotive force on itinerant spins in the metals. This phenomenon should be revealed as a negative drag in an open two-terminal geometry, or equivalently, as a positive magnetoresistance when the terminals are connected in parallel. We propose non-local transport measurements with these salient features as a tool to characterize the phase diagram of insulating chiral magnets.
\end{abstract}

\title{Gyrotropic elastic response of skyrmion crystals to current-induced tensions}

\maketitle

\textit{Introduction}.---Topological solitons in magnetic materials are non-linear excitations with a well-defined energy that behave as particles. Their dynamics can be manipulated by different means, like spin-polarized currents\cite{electrical_manipulation1} or thermal gradients.\cite{thermal_manipulation1} One example of these excitations are magnetic skyrmions, continuous spin textures characterized by an integer charge\cite{Belavin_Polyakov}\begin{align}
\label{eq:topo_charge}
\mathcal{Q}\equiv\frac{1}{4\pi}\int d^2\vec{x}\,\, \bm{n}\cdot\left(\partial_x\bm{n}\times\partial_y\bm{n}\right),
\end{align}
where $\bm{n}$ is a unit vector along the local spin density, $\bm{s}\left(\vec{x}\right)$. This topological charge labels the number of times that the local order parameter wraps the unit sphere in a planar ferromagnet. In the simplest approximation, a rigid skyrmion behaves as a massless particle subjected to a Magnus or gyrotropic force proportional to $\mathcal{Q}$.\cite{Magnus} This force deflects the trajectory with respect to the direction of the driving force, leading to the so-called skyrmion Hall effect.\cite{sky_Hall}

Bogdanov and others\cite{sky_crystal1,sky_crystal2,sky_crystal3} predicted the existence of a crystalline phase in which the skyrmions are spontaneously arranged in a two-dimensional lattice. This phase, stabilized by relativistic interactions in the presence of a magnetic field perpendicular to the magnet, can be visualized as the close packing of individual skyrmions forming a triangular lattice. The skyrmion crystal has been observed in various systems, ranging from itinerant chiral magnets like MnSi\cite{MnSi,MnSi_films} or FeGe\cite{FeGe1,FeGe2} to multiferroic insulating materials like Cu$_2$OSeO$_3$.\cite{ferroic1,ferroic2}

\begin{figure}[t!]
\begin{center}
\hspace{-0.4cm}
\includegraphics[width=0.85\columnwidth]{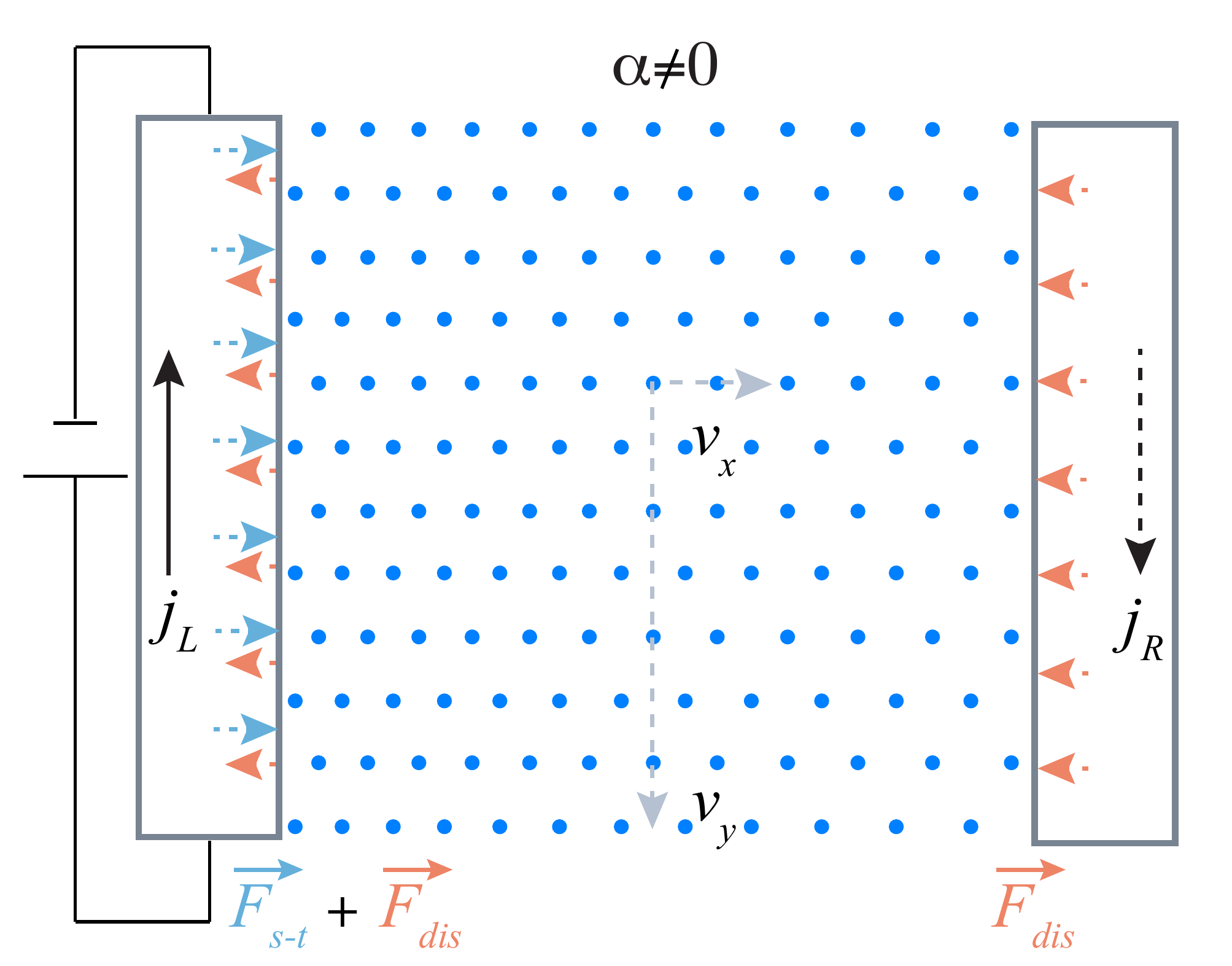}
\caption{Two-terminal geometry considered in the text. A current supplied by an external source in one of the metal contacts exerts a tension $\vec{F}_{\textrm{s-t}}$ over the skyrmion crystal (the blue points represent the center of mass of the skyrmions). In the absence of Gilbert damping $\alpha$, the crystal moves collectively parallel to the electrical current. The damping deflects the trajectory, pushing the skyrmions from one contact to the other and generating an electromotive force in the second terminal. The red arrows represent the viscous force $\vec{F}_{\textrm{dis}}$ due to the enhanced (effective) Gilbert damping near the interface.}
\vspace{-0.5cm} 
\label{fig:open}
\end{center}
\end{figure}

In this Rapid Communication, we study the dynamics of skyrmion crystals in the steady-flow-motion regime. We focus on electrically-insulating thin films, in which the forces are induced by spin-transfer torques\cite{torque,torque2} at the interface with diffusive metals. To that end, we develop a generalized elasticity theory describing the collective dynamics of skyrmions, where the internal stress arises from the exchange interaction between localized spins. The formalism relies on general symmetry arguments and related conservation laws, so it can be extended to different systems, from magnetic bubbles\cite{bubbles} to vortices in layered superconductors.\cite{rev_sc} Our theory provides the basic ingredients for a full-electrical measurement of the skyrmion dynamics that can be used as an alternative to neutron scattering\cite{electrical_manipulation2} and transmission electron microscopy.\cite{thermal_manipulation2}

\textit{Main results}.---We disclose first the general expressions of the elasticity theory that will be applied to the two-terminal geometry depicted in Fig.~\ref{fig:open}. The equation of motion for the displacements of the skyrmions $\vec{u}=\left(u_x,u_y\right)$ within the plane of the magnet reads
\begin{gather}
 \left(\frac{\alpha s}{\Omega}-\frac{4\pi s\mathcal{Q}}{\Omega}\,\hat{z}\times\right)\dot{\vec{u}}=\mu\,\nabla^2\vec{u}+
\left(\lambda+\mu\right)\vec{\nabla}\left(\vec{\nabla}\cdot\vec{u}\right),
\label{eq:elasticity}
\end{gather}
where $s\equiv \left|\bm{s}\left(\vec{x}\right)\right|$ is the saturated spin density and $\Omega$ is the area of the skyrmion lattice unit-cell. The left-hand side of Eq.~\eqref{eq:elasticity} contains a dissipative force proportional to $\alpha$, accounting for Gilbert damping,\cite{Gilbert} and the gyrotropic force arising from the non-trivial topology of the spin texture. The right-hand side corresponds to the internal stress engendered by the displacements of the skyrmions with respect to their equilibrium position. This response (shear and compression) is characterized by two elastic constants, $\mu$, $\lambda$, owing to the hexagonal symmetry of the skyrmion lattice. The order of magnitude is set by $\mathcal{D}^2/\mathcal{A}$ in both cases,\cite{Kwan_etal,Nagaosa,Petrova_Tchernyshyov} where $\mathcal{D}$ and $\mathcal{A}$ are the strength of the Dzyaloshinskii-Moriya coupling and the magnetic stiffness of the film, respectively.

Equation \eqref{eq:elasticity} must be supplemented by boundary conditions reflecting the exchange of energy and linear momentum with the metallic reservoirs, as depicted by the arrows in Fig.~\ref{fig:open}. The current-induced spin torque at the interface with a metal, provided a strong exchange interaction by proximity with the magnet, works in favor of the nucleation of skyrmions,\cite{skyrmions} applying then a tension of the form
\begin{align}
\label{eq:Fst}
\vec{F}_{\textrm{s-t}}=\frac{2\pi\hbar\mathcal{P}\mathcal{Q}\xi}{e\Omega}\,\hat{z}\times\vec{j},
\end{align}
where $\vec{j}$ is the current density in the adjacent metal. The dimensionless parameter $\mathcal{P}$ and the length $\xi$ measure the strength and spatial extension of the proximity effect. Reciprocally, the annihilation of skyrmions at the boundaries generates an electromotive force\begin{align}
\label{eq:pumping}
\vec{\mathcal{E}}_{\textrm{pump}}=\frac{2\pi\hbar\mathcal{P}\mathcal{Q}}{e\Omega}\, \hat{z}\times\dot{\vec{u}}\,|_{b}\,,
\end{align}
where $\dot{\vec{u}}\,|_{b}$ denotes the velocity of skyrmions perpendicular to the interface. The generation of a spin current in the metal dissipates energy and angular momentum from the magnet,\cite{clement_zhang} giving rise to a viscous tension at the interface,
\begin{align}
\label{eq:Fpump}
\vec{F}_{\textrm{dis}}=-\left(\frac{2\pi\hbar}{e\Omega}\right)^2\frac{\xi\,\dot{\vec{u}}\,|_{b}}{\varrho},
\end{align}
where $\varrho$ is the resistivity of the metal. Equations~\eqref{eq:Fst}~and~\eqref{eq:Fpump} define the boundary conditions in the steady state, corresponding to the balance between the applied tensions and the internal stress.

These equations along with Ohm's law in the metals constitute the basic elements of the self-consistent magneto-electric dynamics. Equipped with this formalism, we study the response of two metallic layers connected by an electrically-insulating chiral magnet in the skyrmion-crystal phase. In the open geometry of Fig.~\ref{fig:open}, the electrical current supplied by an external source applies a spin-transfer tension in one of the terminals. In the absence of dissipation, the skyrmion crystal moves perpendicularly to the tension due to the gyrotropic force, parallel to the current. There is no pumped current in the second terminal in that case. In the presence of dissipation, however, viscous forces generate a longitudinal motion, which pumps an electrical current in the right terminal. The effect is characterized by a dimensionless drag coefficient of the form\begin{align}
\label{eq:drag}
\mathcal{C}_d=-\frac{\mathcal{P}^2\,g_R}{g_{\alpha}+g_L+g_R}.
\end{align}
Here $g_{L,R}=\xi\left(\varrho_{L,R}\right) ^{-1}$ are the effective interfacial conductances, whereas $g_{\alpha}$ parametrizes the dissipation of energy by the skyrmion dynamics,
\begin{align}
g_{\alpha}=\frac{s\,e^2\,\Omega\, L\left(\alpha^2+16\pi^2\right)}{4\pi^2\alpha\,\hbar^2}.
\end{align}
Notice that $\left(\alpha^2+16\pi^2\right)/\alpha$ is the medium viscosity for skyrmions with $\mathcal{Q}=\pm 1$. The drag signal decays algebraically with the distance between contacts, $L$, reflecting the conservation of the skyrmion charge.\cite{skyrmions} Taking $s/\hbar\sim 10^{21}$ cm$^{-3}$, $\alpha=0.1$ and a separation of $L=100$ nm, and assuming $\mathcal{P}\sim 1$ over a distance $\xi=1$ nm in Co terminals, we obtain drag signals of $10^{-5}$ in Cu$_2$OSeO$_3$ thin films\cite{ferroic1} with $\Omega\sim 100$ nm$^2$, comparable to recent magnon-drag experiments in yttrium iron garnet.\cite{magnon-drag} Notice that the signal can be further enhanced by increasing the value of $\alpha$ (the signal grows when $\alpha$ increases up to $4\pi$) or $\Omega$, e.g., by introducing appropriate dopants or a heavy-metal substrate. This drag effect and related non-local transport signatures can be used to characterize the phase diagram of insulating chiral magnets.

\textit{Elasticity theory}.--- The starting point of our discussion is the Landau-Lifshitz equations\cite{LL} describing the classical dynamics of the spin-density field, $\bm{s}\left(\vec{x}\right)$. We neglect dissipation for the moment. At temperatures well below the ordering temperature, $T\ll T_c$, the dynamics is generated by the Poisson-brackets algebra,\cite{Haldane} $\left\{s_i\left(\vec{x}\right),s_j\left(\vec{y}\right)\right\}=\epsilon_{ijk}\,s_k\left(\vec{x}\right)\delta\left(\vec{x}-\vec{y}\right)$, 
with the Hamiltonian $\mathcal{U}$ corresponding to the free energy functional of the magnet; the equation of motion is $\dot{\bm{s}}\left(\vec{x}\right)=\left\{\bm{s}\left(\vec{x}\right),\mathcal{U}\right\}=\bm{s}\left(\vec{x}\right)\times\bm{h}\left(\vec{x}\right)$, where $\bm{h}\equiv-\delta_{\bm{s}}\,\mathcal{U}$ is the force conjugate to $\bm{s}\left(\vec{x}\right)$. The dynamics of stable magnetic textures can be described in terms of a set of collective coordinates parametrizing the slow modes of the system.\cite{collective_coordinates} In our case, these soft modes correspond to the center of mass of the skyrmions regarded as rigid textures, provided that the system is translationally invariant, $\partial_i\,\mathcal{U}=0$; we can write in general $\bm{s}\left(\vec{x},t\right)\longrightarrow\bm{s}\left(\vec{x}-\vec{r}_i\left(t\right)\right)\equiv \bm{s}\left[\vec{r}_i\left(t\right)\right]$. In a continuum description, the collective coordinates $\vec{r}_i$ are promoted to a field $\vec{u}\left(\vec{x}\right)$ describing the displacements of the skyrmions with respect to their equilibrium positions in the lattice. These fields verify the relations
\begin{align}
\label{eq:brackets}
\left\{u_i\left(\vec{x}\right),u_j\left(\vec{y}\right)\right\}=\frac{\Omega}{4\pi s\mathcal{Q}}\,\epsilon_{ij}\,\delta\left(\vec{x}-\vec{y}\right),
\end{align}
as it is deduced from the Poisson-brackets algebra.\cite{SM} Notice that Eq.~\eqref{eq:brackets} resembles the Poisson bracket relations between the guiding centers of particles subjected to a magnetic field $\mathcal{B}\propto 4\pi\mathcal{Q}$.\cite{momentum_topo} The equation of motion reads
\begin{align}
\label{eq:Hamilton}
& \dot{\vec{u}}\left(\vec{x}\right)=\left\{\vec{u}\left(\vec{x}\right),\mathcal{U}\right\}=\frac{\Omega}{4\pi s\mathcal{Q}}\,\hat{z}\times \vec{f}\left(\vec{x}\right),
\end{align}
with $\vec{f}\equiv -\delta_{\vec{u}}\,\mathcal{U}$. In this approximation, we are neglecting the effect of hard modes of the magnetization dynamics, which can be introduced as an inertial term in the equation of motion. This mass accounts for the deformation of the moving skyrmion compared with the static solution and alters the lattice dynamics at high frequencies,\cite{Petrova_Tchernyshyov} but it is inconsequential for the steady-state motion\cite{collective_coordinates} and will be ignored for the rest of the discussion.

The field conjugate to $u_i\left(\vec{x}\right)$ in this approximation is $\pi_i\left(\vec{x}\right)=4\pi s\,\mathcal{Q}\,\epsilon_{ij}\,u_j\left(\vec{x}\right)/\Omega$; indeed, these field variables obey the canonical relation $\left\{u_i\left(\vec{x}\right),\pi_j\left(\vec{y}\right)\right\}=\delta_{ij}\,\delta\,\left(\vec{x}-\vec{y}\right)$. Thus, the total linear momentum of the crystal reads $P_i\equiv\int d\vec{x}\,\pi_i\left(\vec{x}\right)$. 
This is a proper momentum functional since $P_i$ is the generator of spatial translations; in particular, for the free energy we have $\dot{P}_i=\left\{P_i,\mathcal{U}\right\}=\partial_i\,\mathcal{U}$, and we obtain the canonical conservation law for translational invariance. The latter has consequences in the form of the conservative forces in Eq.~\eqref{eq:Hamilton}. The condition $\dot{P}_i=0$ implies that the force density must be written as the divergence of a tensor of rank two, $f_i\left(\vec{x}\right)=\partial_j\sigma_{ij}\left(\vec{x}\right)$; $\sigma_{ij}\left(\vec{x}\right)$ is the stress tensor field. Invariance under rotations about the $\hat{z}$-axis --normal to the plane of the film-- implies that the stress tensor is symmetric, $\sigma_{ij}=\sigma_{ji}$. Hence, the work density carried out by the internal forces of the magnet due to a change in the position of the skyrmions can be evaluated as $\delta \bar{W}=-\sigma_{ij}\,\delta u_{ij}$, where $u_{ij}\equiv\frac{1}{2}\left(\partial_i u_j+\partial_j  u_i\right)$ is the strain tensor field within the plane of the film. At constant temperature we have $\delta\, \bar{\mathcal{U}}=\sigma_{ij}\,\delta u_{ij}$, and we deduce the the usual thermodynamic relation $\sigma_{ij}=\left(\partial\, \bar{\mathcal{U}}/\partial\, u_{ij}\right)_{T}$.\cite{elasticity_theory} The free energy can be written then as a functional of $u_{ij}$; to the lowest order,\cite{foot_thermo} we have\begin{align}
\label{eq:U}
\mathcal{U}=\frac{1}{2}\int d\vec{x}\,\left[\lambda\left(u_{ii}\right)^2+2\,\mu\, u_{ij}u_{ij}\right],
\end{align}
and hence the stress tensor reads
\begin{align}
\sigma_{ij}=\lambda\,u_{kk}\,\delta_{ij}+2\,\mu\,u_{ij}.
\end{align} 
Equation~\eqref{eq:U} describes an isotropic crystal, which applies in particular to the hexagonal skyrmion lattice. For Cu$_2$OSeO$_3$ films we estimate $\lambda,\mu\sim\mu$eV/nm$^3$.\cite{ferroic1,ferroic2}


Dissipation can be introduced phenomenologically in the same spirit as Gilbert damping\cite{Gilbert} by means of a Rayleigh functional of the form $\mathcal{R}=\alpha s\int d\vec{x}\,\dot{\vec{u}}\,^2/\,2\Omega$. The dimensionless coefficient $\alpha$ corresponds to the Gilbert damping constant multiplied by a geometric factor depending on the profile of the texture. Both conservative, $\partial_j\sigma_{ij}$, and dissipative forces $-\delta_{\dot{\vec{u}}}\,\mathcal{R}=-\alpha\, s\,\dot{\vec{u}}/\,\Omega$ enter in Eq.~\eqref{eq:Hamilton}, leading to the final result in Eq.~\eqref{eq:elasticity}

\begin{figure}[t!]
\begin{center}
\hspace{-0.4cm}
\includegraphics[width=0.8\columnwidth]{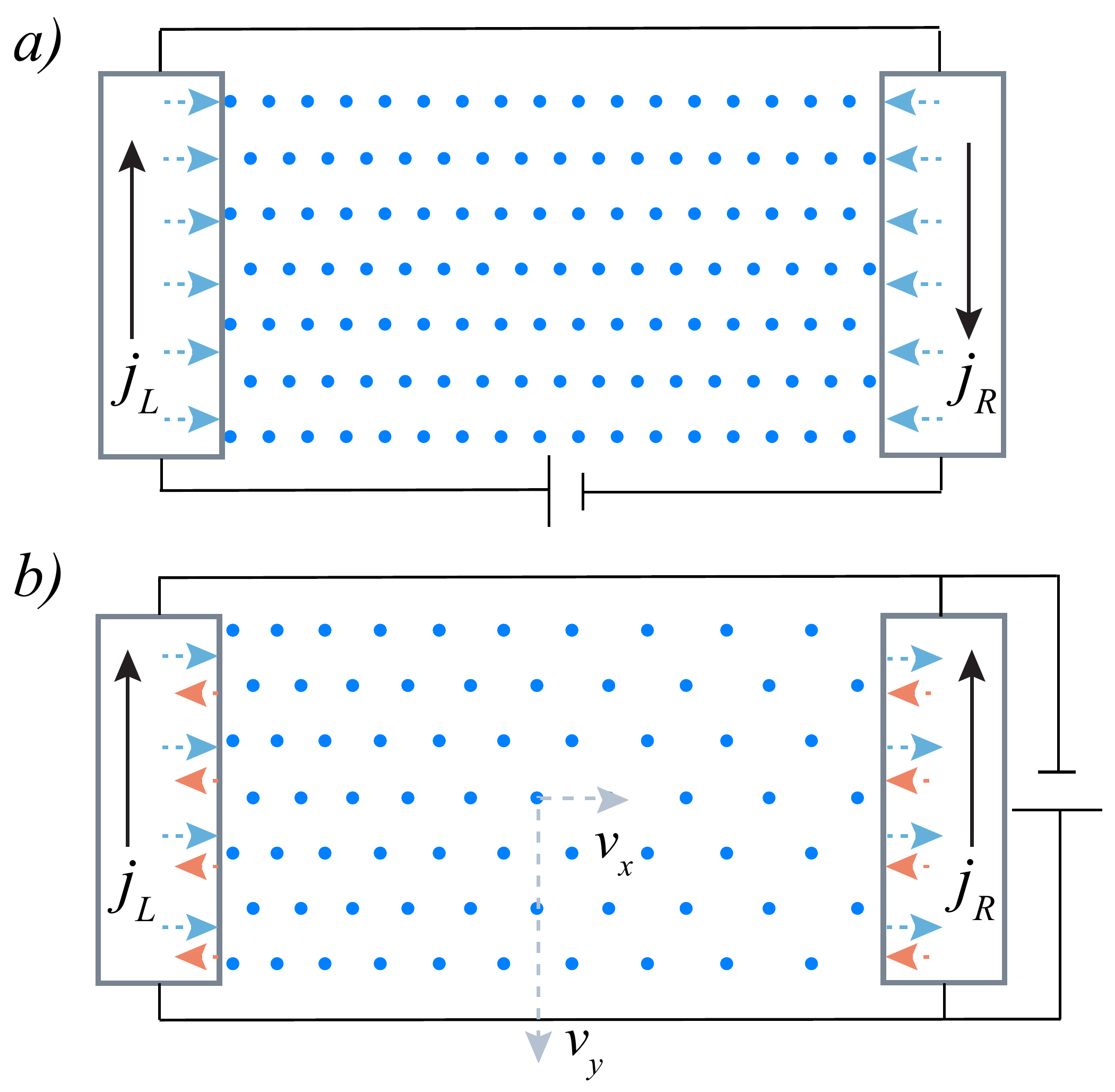}
\caption{Closed circuit in series (a) and parallel (b) configurations. In the former case the skyrmion crystal is compressed but remains static, whereas in the parallel circuit the dynamics of the skyrmions enhances the resistivity of the external circuit, reflecting the negative drag in the open geometry.}
\vspace{-0.8cm} 
\label{fig:closed}
\end{center}
\end{figure}

\textit{Driving forces and pumping}.---We assume a strong exchange interaction between the itinerant spins in the metal and the localized spins in the magnet. This proximity effect extends over a certain length $\xi$ in the metal. In the magnet, the Landau-Lifshitz equation at the interface must be supplemented with the non-equilibrium torque exerted by a spin-polarized electrical current, $\dot{\bm{s}}\left(\vec{x}\right)=\left\{\bm{s}\left(\vec{x}\right),\mathcal{U}\right\}\,+\,\bm{\tau}_{\textrm{s-t}}$, with $\bm{\tau}_{\textrm{s-t}}=\frac{\hbar}{2e}\mathcal{P}\,\vec{j}\cdot\vec{\nabla}\,\bm{n}$ in the adiabatic limit,\cite{pumping} i.e., to the lowest order in spatial gradients of the magnetization and neglecting spin relaxation. Here $\mathcal{P}$ measures the spin polarization of the current in the adjacent metal. This torque arises form the exchange of linear momentum between the metal and the magnet and therefore applies a force on the skyrmion crystal. This tension can be computed from the work-power exerted by the spin-transfer torque, $\dot{\bm{n}}\cdot\left(\bm{\tau}_{\textrm{s-t}}\times\bm{n}\right)$, integrated over the length $\xi$. In our collective field approach we have $\dot{\bm{n}}\approx-\,\dot{\vec{u}}\cdot\vec{\nabla}\bm{n}$, and the resulting expression can be related to the skyrmion charge density since $\bm{n}\cdot\left(\partial_i\bm{n}\times\partial_j\bm{n}\right)\approx 4\pi\mathcal{Q}\epsilon_{ij}/\Omega$. We obtain the expression in Eq.~\eqref{eq:Fst} by identifying the power-work with $\dot{\vec{u}}\cdot\vec{F}_{\textrm{s-t}}$.

Reciprocally, the skyrmion dynamics at the interface induces an electromotive force in the metal of the form\cite{pumping}\begin{align}
\vec{\mathcal{E}}_{\textrm{pump}}=\frac{\hbar}{2e}\mathcal{P}\,\bm{n}\cdot\left(\vec{\nabla}\bm{n}\times\dot{\bm{n}}\right).
\end{align}
Eq.~\eqref{eq:pumping} is directly derived from this equation in the collective field approximation. This pumped force results from the spin-dependent electric field generated in the metal due to the accumulation of Berry phases\cite{Berry} by itinerant spins adiabatically following the exchange field. Thus, as long as the adiabatic approximation holds a spin current is also generated in the metal,\cite{clement_zhang}\begin{align}
\label{eq:js}
\vec{\bm{j}}_s=\frac{\hbar^2}{4\, e^2\varrho}\,\vec{\nabla}\bm{n}\times\dot{\bm{n}}\approx\frac{\pi\,\hbar^2\mathcal{Q}}{e^2\varrho\,\Omega}\left(\hat{z}\times\dot{\vec{u}}\right) \bm{n}.
\end{align}
The depletion of spin angular momentum exerts a dissipative torque on the magnetization. In the absence of spin relaxation we have $\bm{\tau}_{\textrm{dis}}=\vec{\nabla}\cdot\vec{\bm{j}}_s$,\cite{clement_zhang} leading to the viscous tension in Eq.~\eqref{eq:Fpump}.\cite{foot_Fdis} Notice that this expression breaks explicitly the macroscopic time-reversal symmetry of the skyrmion dynamics. The force is perpendicular to the electron flow and therefore normal to the boundaries of the magnet. A correction to the spin-motive force ($\propto\beta\, \dot{\bm{n}}\cdot\vec{\nabla}\bm{n}$) due to spin relaxation may generate an additional shear viscosity, tilting the skyrmion crystal with respect to its equilibrium configuration.\cite{torque2} We neglect this contribution provided a strong exchange coupling at the interface ($\beta\ll 1$).

\textit{Spin-transfer drag}.--- We address now the magneto-electric dynamics in the open geometry depicted in Fig.~\ref{fig:open}. We assume that the magnetic field stabilizing the skyrmion-crystal phase points in the positive $\hat{z}$-axis, therefore $\mathcal{Q}=-1$; we also assume translational invariance along the interface. Above certain current-density threshold determined by pinning forces, the spin-transfer tension translates the skyrmion lattice with velocity $\vec{v}=\left(v_x,v_y\right)$. In the steady state the internal stress does not depend on time, so the skyrmion displacements with respect to the co-moving frame with the lattice are stationary, $\vec{u}\left(x\right)$. The bulk dynamics expressed in Eq.~\eqref{eq:elasticity} relates the skyrmion-lattice velocity with the internal stress. The latter is determined by the boundary conditions $\vec{F}_{\textrm{s-t}}+\vec{F}_{\textrm{dis}}=\pm\,\sigma_{xx}\,\hat{x}$, where the $+$ ($-$) sign applies to the right (left) terminal. The velocity in linear response\cite{foot_convective} reads
\begin{align}
v_x=-\frac{\alpha}{4\pi}\, v_y=\frac{e\,\Omega\,\xi\,\mathcal{P}\,j_L}{2\pi\hbar\left(g_{\alpha}+g_L+g_R\right)}.
\end{align}
Notice that in the absence of dissipation the skyrmion crystal moves parallel to the current and there is no pumping in the second terminal. From this solution and Eq.~\eqref{eq:pumping} we determine the current density pumped in the second terminal, $\vec{j}_R=\varrho_R^{-1} \vec{\mathcal{E}}_{\textrm{pump}}$; we obtain then the dimensionless drag coefficient $\mathcal{C}_{d}\equiv j_R/j_L$ in Eq.~\eqref{eq:drag}.

\begin{figure}[t!]
\begin{center}
\hspace{-0.4cm}
\includegraphics[width=0.85\columnwidth]{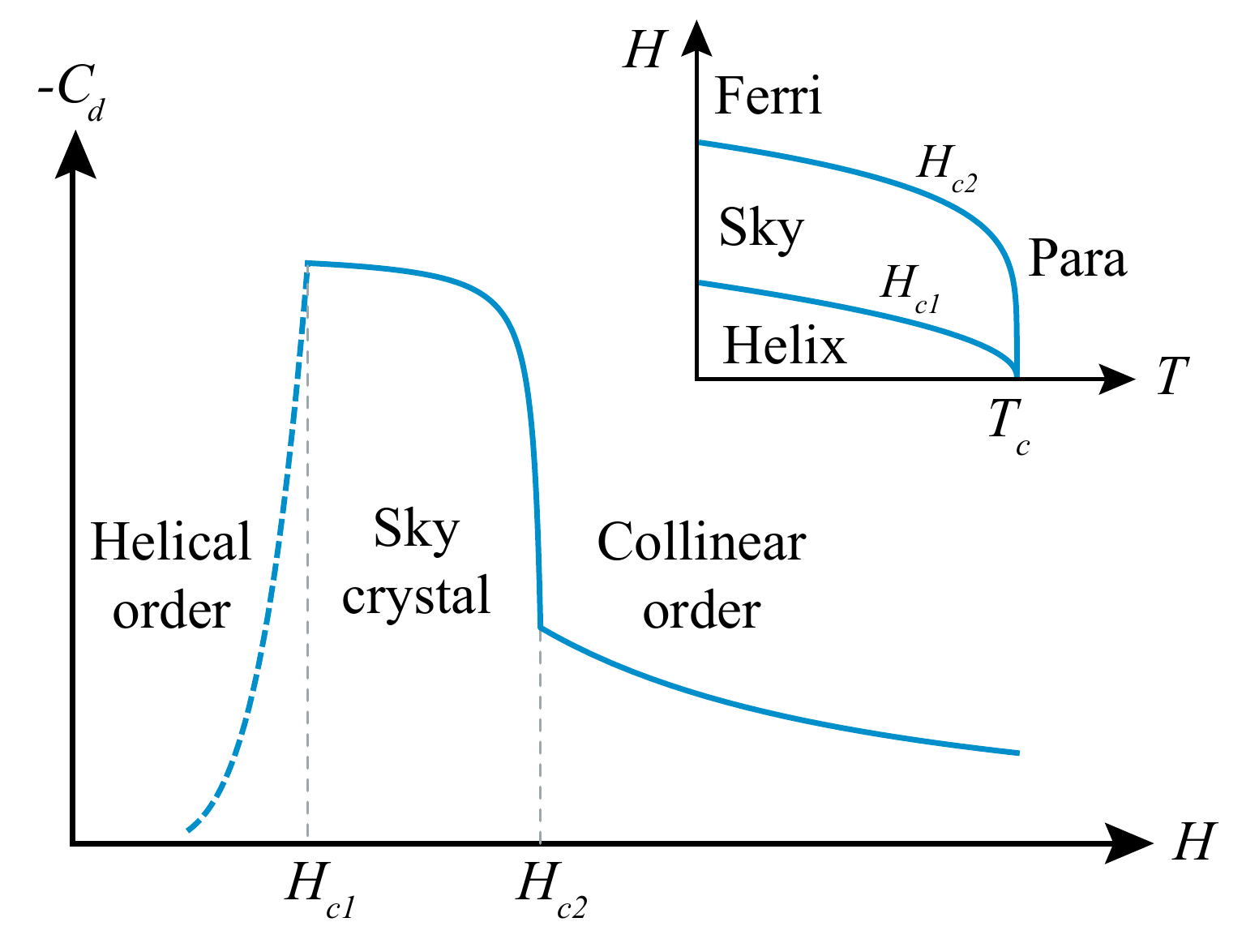}
\caption{Drag coefficient as a function of applied magnetic field $H$ at temperatures well below $T_c$. The inset shows the schematic magnetic phase diagram in thin films of Cu$_2$OSeO$_3$ as deduced from transmission electron microscopy and magnetic susceptibility measurements.\cite{ferroic1}}
\vspace{-0.5cm} 
\label{fig:drag}
\end{center}
\end{figure}

\textit{Non-local magnetoresistance}.--- When the two terminals are connected as in Fig.~\ref{fig:closed}, the drag effect results in a non-local magnetoresistance depending on the configuration of the external circuit. We assume that both terminals are identical for simplicity, $\varrho\equiv\varrho_L=\varrho_R$. Solving the magneto-electric dynamics in this case, taking into account now the additional spin-transfer tension in the right contact, leads to the pumping electromotive force\begin{align}
\vec{\mathcal{E}}_{\textrm{pump}}=-\frac{\mathcal{P}^2\xi}{g_{\alpha}+2g_i}\left(\vec{j}_L+\vec{j}_R\right).
\end{align} In the series configuration, Fig.~\ref{fig:closed}~a), we have $\vec{j}_{L}=-\vec{j}_R$ and therefore no pumping; the spin-transfer tensions are applied in opposite direction at each terminal, the skyrmion crystal is then compressed but remains static. In the parallel circuit, Fig.~\ref{fig:closed}~b), we have $\vec{j}_L=\vec{j}_R\equiv \vec{j}$ and therefore $\vec{\mathcal{E}}_{\textrm{pump}}=2\,\varrho\,\mathcal{C}_d\,\vec{j}$. This modifies Ohm's law in the metals as $\varrho\,\vec{j}=\vec{\mathcal{E}}+\vec{\mathcal{E}}_{\textrm{pump}}$, where $\vec{\mathcal{E}}$ is the electromotive force supplied by the external source. Thus, the effective resistivity $\varrho'$ of the circuit, $\varrho'\,\vec{j}=\vec{\mathcal{E}}$, acquires an additional non-local correction, $\varrho'=\varrho+\varrho_m$; the magnetoresistance reads\begin{align}
\varrho_m=-2\,\mathcal{C}_d\,\varrho=\frac{\mathcal{P}^2\xi}{g_{\alpha}/2+g_i}.
\end{align}

\textit{Discussion}.--- It is worth comparing Eq.~\eqref{eq:drag} with the drag coefficient obtained for a gas of metastable skyrmions.\cite{skyrmions} For a large separation between contacts, $g_{\alpha}\gg g_{L,R}$, the drag coefficient adopts the following general expression, valid in both phases:
\begin{align}
\mathcal{C}_d\approx-\frac{\mu_{\textrm{sky}}\,\xi\,\rho_{\textrm{sky}}}{\varrho}\left(\frac{2\pi\hbar\mathcal{P}}{e}\right)^2\frac{d}{L}.
\end{align}
Here $d$ is the thickness of the film, $\rho_{\textrm{sky}}$ is the density of skyrmions at equilibrium, and $\mu_{\textrm{sky}}=\frac{\alpha}{s\,d\,\left(16\pi^2+\alpha^2\right)}$ is their mobility. The expected behavior of the drag coefficient as a function of the applied magnetic field is shown in Fig.~\ref{fig:drag} for thin films of Cu$_2$OSeO$_3$, whose phenomenological phase diagram is depicted in the inset. At large magnetic fields the ground state is uniformly ordered. The drag signal is driven by the Brownian motion of thermally activated skyrmions, $\rho_{\textrm{sky}}\propto e^{-E\left(H\right)/k_BT}$, decaying exponentially with $H$. The distance between isolated skyrmions decreases as the magnetic field approches $H_{c2}$, the critical field at which the skyrmions forms a regular lattice, $\rho_{\textrm{sky}}\approx\Omega^{-1}$. The unit-cell area remains approximately unchanged at lower fields. Notice that this transition from the skyrmion-crystal side is likely to be anticipated by the lattice melting due to thermal fluctuations, similarly to the mixed state of layered type II superconductors;\cite{typeII} in that regard, the critical line $H_{c2}\left(T\right)$ should be taken merely as a crossover. The core of the skyrmions increases monotonically as $H$ decreases\cite{Bogdanov_Hubert} and the system enters into the helically ordered phase at $H_{c1}$. The proposed non-local transport measurements can provide some insights about the nature of this phase transition, in particular the role of topological defects such as disclinations\cite{disclinations} and meron-like excitations\cite{Ezawa} that couple to the electrical currents as expected from our theory. Finally, the spin-transfer drag effect should be detected only within a current-density threshold $j_{c1}<j<j_{c2}$. The lower critical current is determined by pinning forces, of the order of $j_{c1}\sim 10^6$ A$/$m$^{2}$,\cite{electrical_manipulation2} whereas the upper critical current is related to the breakdown tension above which the skyrmions at the left terminal overlap, $j_{c2}\approx e\,\Omega\left(\lambda+2\mu\right)/2\pi\hbar\mathcal{P}\xi \sim 10^{10}$ A$/$m$^{2}$.

In summary, we have constructed the elasticity theory and irreversible thermodynamics of skyrmion crystals coupled to electrical currents in adjacent metals. The theory has been employed to study the long-range drag signal and related magnetoresistance signatures induced by a steady-state skyrmion motion between two metallic terminals. These ideas can be tested in thin films of Cu$_2$OSeO$_3$, provided that the crystal phase extends in a wide range of temperatures and magnetic fields.


This work was supported by the U.S. Department of Energy, Office of Basic Energy Sciences, Division of Materials Science and Engineering under awards DE-SC0012190 (UCLA) and DE-FG02-08ER46544 (JHU).

\clearpage

\section*{Supplementary material}

Eq.~(8) of the main text can be understood as a continuum version of the Poisson brackets between the coordinates of the center of mass of a single skyrmion,\cite{momentum_topo}\begin{align}
\left\{r_i^x,r_j^y \right\}=\frac{\delta_{ij}}{4\pi s\mathcal{Q}}.
\end{align}
Here the indices $i$, $j$ label de positions in the skyrmion crystal. In the continuum version of this equation, the coordinates $\vec{r}_i$ are promoted to fields and the Kronecker delta is substituted by a Dirac delta, regularized by the area of the unit cell, $\Omega$. For the rest of this Supplementary Material, we provide a more rigorous derivation starting from the Poisson-brackets algebra for the spin-density fields.

Let $f$ and $g$ be two dynamical variables, i.e., functionals of the magnetic texture $\bm{s}\left(\vec{x}\right)$. The Poisson brackets between them are given by (we adopt the Einstein summation convention from now on)
\begin{align}
&\left\{f,g\right\}=\int d\vec{x}\int d\vec{y}\,\frac{\delta f}{\delta s_i\left(\vec{x}\right)}\frac{\delta g}{\delta s_j\left(\vec{y}\right)}\left\{s_i\left(\vec{x}\right),s_j\left(\vec{y}\right)\right\}
\nonumber\\
&=\int d\vec{x}\,\bm{s}\left(\vec{x}\right)\cdot\left(\frac{\delta f}{\delta \bm{s}\left(\vec{x}\right)}\times\frac{\delta g}{\delta \bm{s}\left(\vec{x}\right)}\right).
\end{align}
Given a magnetic texture describing the skyrmion crystal, the displacement fields are univocally defined. Thus, the Poisson brackets in Eq.~(8) of the main text reads in general
\begin{align}
\left\{u_i\left(\vec{x}\right),u_j\left(\vec{y}\right)\right\}=\int d\vec{z}\,\bm{s}\left(\vec{z}\right)\cdot\left(\frac{\delta u_i\left(\vec{x}\right)}{\delta \bm{s}\left(\vec{z}\right)}
\times\frac{\delta u_j\left(\vec{y}\right)}{\delta \bm{s}\left(\vec{z}\right)}\right).
\end{align}

Let us define the gyrotropic tensor field as
\begin{align}
\label{eq:G}
G_{ij}\left(\vec{x},\vec{y}\right)\equiv\int d\vec{z}\,\bm{s}\left(\vec{z}\right)\cdot\left(\frac{\delta \bm{s}\left(\vec{z}\right)}{\delta u_i\left(\vec{x}\right)}\times\frac{\delta \bm{s}\left(\vec{z}\right)}{\delta u_j\left(\vec{y}\right)}\right).
\end{align}
This tensor is the inverse of the reduced symplectic form in the displacement-field variables. After some algebra, we arrive at the identity: \begin{widetext}
\begin{align}
\int d\vec{z}\, G_{ik}\left(\vec{x},\vec{z}\right)\left\{u_k\left(\vec{z}\right),u_j\left(\vec{y}\right)\right\}=-s^2\int d \vec{z}\,\frac{\delta u_j\left(\vec{y}\right)}{\delta \bm{s}\left(\vec{z}\right)}\cdot\frac{\delta \bm{s}\left(\vec{z}\right)}{\delta u_i\left(\vec{x}\right)}+\int d\vec{z}\,\left(\bm{s}\left(\vec{z}\right)\cdot\frac{\delta \bm{s}\left(z\right)}{\delta u_i\left(\vec{x}\right)}\right)\left(\bm{s}\left(\vec{z}\right)\cdot\frac{\delta u_j\left(\vec{y}\right)}{\delta \bm{s}\left(\vec{z}\right)}\right).
\end{align}
\end{widetext}
Notice that the first pair of parentheses in the second term on the right-hand side is proportional to $\delta \left(\left|\bm{s}\left(\vec{x}\right)\right|^2\right)=\delta s^2=0$. Therefore, the second term is identically $0$. For the first term, we have\begin{align}
\int d \vec{z}\,\frac{\delta u_j\left(\vec{y}\right)}{\delta \bm{s}\left(\vec{z}\right)}\cdot\frac{\delta \bm{s}\left(\vec{z}\right)}{\delta u_i\left(\vec{x}\right)}=\frac{\delta u_j\left(\vec{y}\right)}{\delta u_i\left(\vec{x}\right)}=\delta_{ij}\delta\left(\vec{x}-\vec{y}\right),
\end{align}
and hence
\begin{align}
\int d\vec{z}\, G_{ik}\left(\vec{x},\vec{z}\right)\left\{u_k\left(\vec{z}\right),u_j\left(\vec{y}\right)\right\}=-s^2\delta_{ij}\delta\left(\vec{x}-\vec{y}\right).
\end{align}
We can write then
\begin{align}
\label{eq:intermediate}
\left\{u_i\left(\vec{x}\right),u_j\left(\vec{y}\right)\right\}=-s^2\,G_{ij}^{-1}\left(\vec{x},\vec{y}\right).
\end{align}

Within the collective field-coordinates approach, we have the identification:
\begin{align}
\label{eq:ansatz}
\bm{s}\left(\vec{x}\right)\longrightarrow\bm{s}\left(\vec{x}-\vec{u}\left(\vec{x}\right)\right)\equiv\bm{s}\left[\vec{u}\left(\vec{x}\right)\right].
\end{align}
Then, the functional derivatives in the definition of $G_{ij}\left(\vec{x},\vec{y}\right)$ can be approximated as
\begin{align}
\label{eq:continuum}
\frac{\delta \bm{s}\left(\vec{x}\right)}{\delta u_i\left(\vec{y}\right)}\approx-\partial_i\bm{s}\left(\vec{x}\right)\,\delta\left(\vec{x}-\vec{y}\right).
\end{align}
By plugging these expression into the definition of Eq.~\eqref{eq:G} we obtain
\begin{align}
&G_{ij}\left(\vec{x},\vec{y}\right)\approx\bm{s}\left(\vec{x}\right)\cdot\left[\partial_i\bm{s}\left(\vec{x}\right)\times\partial_j\bm{s}\left(\vec{x}\right)\right]\delta\left(\vec{x}-\vec{y}\right)
\nonumber\\
&\approx\frac{4\pi\,s^3\mathcal{Q}}{\Omega}\,\epsilon_{ij}\,\delta\left(\vec{x}-\vec{y}\right).
\end{align}
The last approximation corresponds to a continuum description of the skyrmion lattice, in which the the skyrmion-charge density is approximated by $\mathcal{Q}/\Omega$. Eq.~\eqref{eq:intermediate} reduces to
\begin{align}
\label{eq:brackets}
\left\{u_i\left(\vec{x}\right),u_j\left(\vec{y}\right)\right\}=\frac{\Omega}{4\pi s\mathcal{Q}}\,\epsilon_{ij}\,\delta\left(\vec{x}-\vec{y}\right),
\end{align}
which is the expression in Eq.~(8) of the main text. 


\begin{thebibliography}{99}
\bibitem{foot_Fdis} When $\vec{j}$ is fixed, Eq.~\eqref{eq:js} must be corrected by a factor $\left(1-\mathcal{P}^2\right)$. This, along with geometric prefactors arising from the finite spin diffusion lengths in the metal or the deformations of the skyrmion texture, can be incorporated as a phenomenological parameter of order $1$ in Eq.~\eqref{eq:Fpump} that we omit throughout the Letter.
\bibitem{foot_convective} In Eq.~\eqref{eq:elasticity} we must take $\dot{\vec{u}}=\vec{v}+(\vec{v}\cdot\vec{\nabla})\vec{u}$, corresponding to the skyrmion velocity with respect to the static background. The convective correction can be neglected if $\left|\nabla\vec{u}\right|\ll1$, which is ensured if $j\ll j_{c2}$.
\end{thebibliography}
\end{document}